\documentclass[reprint, superscriptaddress,amsmath,amssymb,
 aps,prd,floatfix,nofootinbib
]{revtex4-2}
\usepackage{graphicx}
\usepackage{comment}
\usepackage{orcidlink}
\usepackage{hyperref}  
\usepackage{dcolumn}
\usepackage{bm}
\usepackage{xcolor}
\usepackage{capt-of}

\begin{document}

\preprint{APS/123-QED}

\title{\textbf{Higher-order statistics of the stochastic gravitational wave background from supermassive black hole binaries} 
}

\author{Hinano Hisamatsu\orcidlink{0009-0004-9045-2517}}
\email{h.hisamatsu@chiba-u.jp}
\affiliation{Department of Physics, Graduate School of Science, Chiba University, Chiba 263-8522, Japan}

\author{Koutarou Kyutoku \orcidlink{0000-0003-3179-5216}}
\email{kyutoku@chiba-u.jp}
\affiliation{Department of Physics, Graduate School of Science, Chiba University, Chiba 263-8522, Japan}
\affiliation{Interdisciplinary Theoretical and Mathematical Sciences Program (iTHEMS), RIKEN, Wako, Saitama 351-0198, Japan}

\date{\today}

\begin{abstract}
Recent progress in gravitational wave observations has positioned pulsar timing arrays as a key tool for detecting the stochastic gravitational wave background in the nanohertz band. It is widely believed that this background is primarily attributed to the cosmic ensemble of inspiraling supermassive black hole binaries. While traditional analyses have predominantly focused on the spectral amplitude and frequency dependence of the gravitational wave background, higher-order statistics such as variance, skewness, and excess kurtosis could potentially be useful for extracting further physical information. 
However, these statistical moments are known to diverge when the redshift integration is extended down to $z=0$. 
In this study, we propose a strategy to resolve this issue by introducing a physically motivated lower integration limit, $z_\mathrm{min}$, defined by the sensitivity for detecting individual sources. Since higher-order statistics are primarily determined by local sources, we may adopt the lowest-order approximation with respect to redshift in their computations. 
Under this approximation, we demonstrate that all higher-order statistics beyond the expectation value depend on the mass function only through a weighted average of the chirp mass, $\langle \mathcal{M}^{10/3} \rangle$, irrespective of the redshift evolution model. We show that the ratio of the variance to the expectation value provides information on $\langle \mathcal{M}^{10/3} \rangle / \langle \mathcal{M}^{5/3} \rangle$ independently of the total number of mergers. We also find a consistency relation between the excess kurtosis and the squared skewness, paving the way for testing the binary-origin hypothesis of the gravitational wave background. Our findings demonstrate that higher-order statistics provide a new window for interpreting the gravitational wave background, offering a methodology to break existing degeneracies and refine our understanding of the mass function.

\end{abstract}

\maketitle

\section{Introduction\label{sec:intro}}

Recent pulsar timing array (PTA) observations have reported the existence of a stochastic gravitational wave background at $2$ to $4\sigma$ levels \cite{agazie2023c,reardon2023,antoniadis2023,arzoumanian2023, miles2024}. The primary candidate for its origin is widely regarded as astrophysical sources, specifically supermassive black hole binaries (SMBHBs). Because no decisive conclusion can be drawn at this stage, interpretations involving cosmological sources like inflation, phase transitions, and topological defects (see, e.g., Refs.~\cite{caprini2018, afzal2023} for reviews) have also gained significant attention and are extensively discussed.

The nature of SMBHBs as well as their formation and evolution processes is yet to be understood. A significant theoretical challenge is the ``final parsec problem'' \cite{1980Natur.287..307B}, which refers to the difficulty of identifying physical processes that can extract sufficient angular momentum from a binary to reduce its orbital separation to the regime where gravitational wave emission leads to coalescence within the age of the Universe. Accurate characterization of the SMBHB population by the gravitational wave background may provide us with a clue to solving this long-standing problem.

For decades, various theoretical studies have predicted the gravitational wave background signal \cite{rajagopal1995,wyithe2003, jaffe2003, sesana2004,sesana2013a,kelley2017a,volonteri2020,Chen_2025}. However, current PTA observations suggest a tension between these theoretical predictions and the actual spectral amplitude \cite{agazie2023b, sato-polito2023, Calz__2024}. Thus, it is of paramount importance to extract detailed information about the source population from observed data to refine the theoretical assumptions. Unfortunately, analyses that focus exclusively on the amplitude and frequency dependence of the spectrum cannot break the inherent parameter degeneracies of the SMBHB population. For instance, this information does not enable us to distinguish whether the background is produced by a vast population of lower-mass binaries or a small number of exceptionally massive binaries.

\begin{figure}[tbp]
    \includegraphics[width=0.9\linewidth]{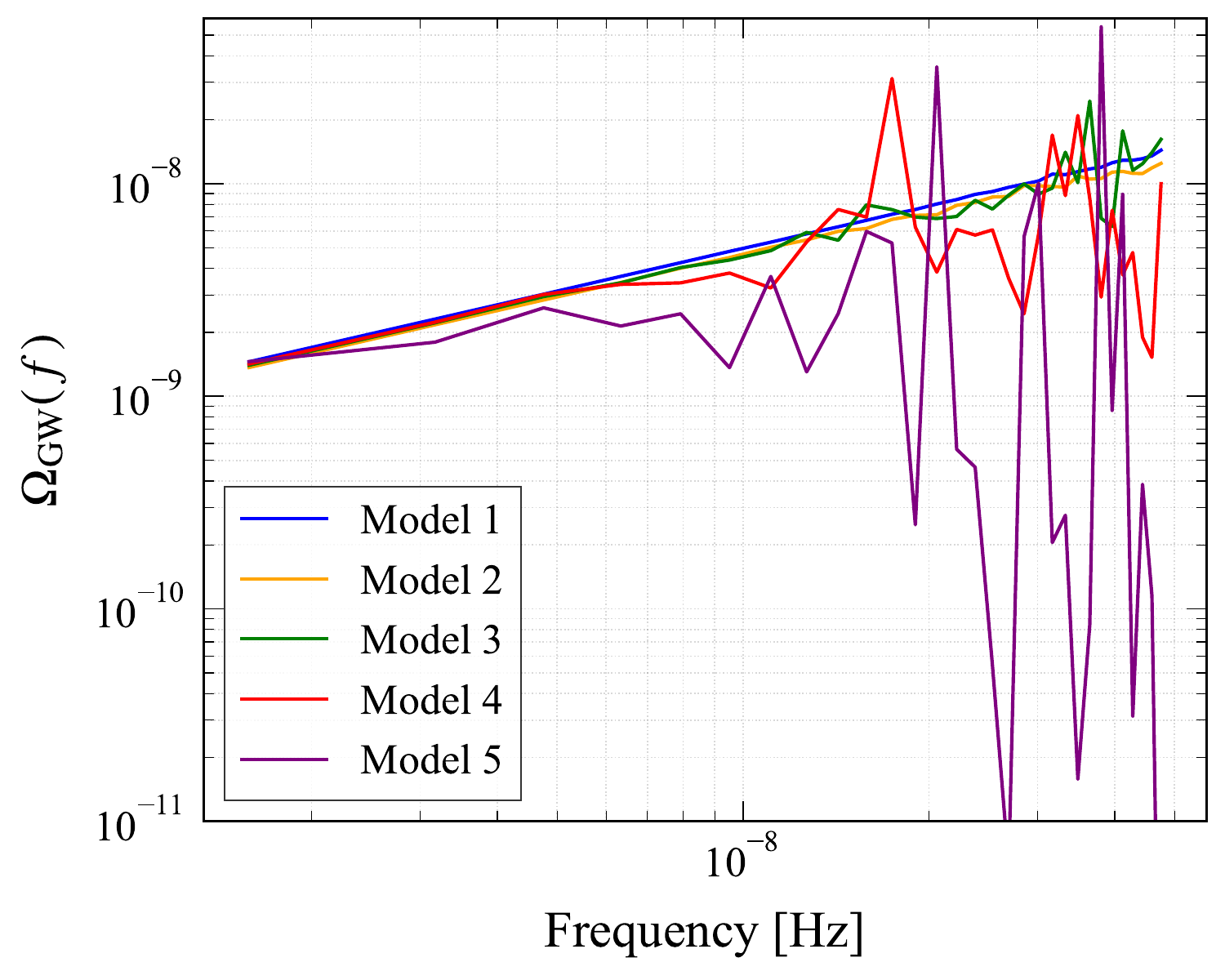}
    \caption{Energy density spectrum calculated via Monte Carlo simulations. The expectation value, or mean, of $\Omega_{\mathrm {GW}}(f)$ is normalized to the same value for comparison. Model 1 represents a population consisting of numerous low-mass sources, while Model 5 is dominated by a small number of high-mass sources. The SMBHB population model and its parameters are adopted from Ref.~\cite{sato-polito2025}. Specifically, the distribution function is given by Eq.~\eqref{presssche} and the parameters for each model are shown in Table~\ref{Modelparam}. Models dominated by a small number of high-mass binaries, such as Model 5, exhibit spiky spectra due to shot noise.}
    \label{fig:fluctuation}
\end{figure}

If the gravitational wave background is indeed composed of emission from SMBHBs, the source population is discrete and finite \cite{agazie2025}. This discreteness implies that the energy density spectrum is not perfectly smooth but exhibits fluctuations, as illustrated in Fig.~\ref{fig:fluctuation}. These fluctuations, known as shot noise, depend on whether binaries are present in specific frequency bands or not. Thus, we expect that the higher-order statistics can provide additional information about the SMBHB population through quantitative characterization of shot noise arising from the finiteness of the sources.

The non-Gaussian nature of the gravitational wave background has been a subject of active research \cite{xue2025, lamb2024, bernardo2025, kuntz2026lookingnongaussianitypulsartiming} (see also Ref.~\cite{Ginat_2024} for studies focusing on ground-based detectors). This non-Gaussianity originates from the discrete nature of the underlying source population and is characterized mainly by the strong ends of the signal distribution \cite{ellis2023, ellis2024, sato-polito2024, raidal2026}. For example, it has been understood how the SMBHB population formally characterizes the magnitude and frequency dependence of higher-order statistics \cite{lamb2024}. However, the integrals required to derive variance and higher-order moments diverge when the source distribution is considered down to $z=0$, because the non-negligible probability of finding exceptionally bright, nearby sources dominates the statistics, thereby obscuring physically meaningful information \cite{jaffe2003}. Accordingly, recent studies have typically focused on the characteristic generating function \cite{xue2025} or the probability density function \cite{ali-haimoud2026}.

We argue that extremely high-amplitude sources responsible for this divergence should not be treated as part of the ``background'' in realistic situations. Instead, they should be identified as individual, resolvable sources and removed from the stochastic background \citep{rosado2015, arzoumanian2023,agazie2023}. Based on this observational perspective, we propose resolving the divergence by introducing physical boundaries that naturally separate resolvable individual sources from the genuine stochastic background. Specifically, this boundary is defined by the amplitude threshold for individual sources and serves as a chirp-mass- and frequency-dependent redshift lower limit, $z_\mathrm{min}(\mathcal{M},f)$. Our approach offers a model-independent alternative to that based on the characteristic generating function or probability density function \cite{xue2025, sato-polito2024, ali-haimoud2026}.

Crucially, the higher-order statistics defined in this study can serve as a tool for breaking the degeneracy between the merger rate and the mass function of SMBHBs, which is inseparable in the expectation value of the gravitational wave background. The ratio of the variance to the expectation value allows us to extract information about the mass function in a robust manner. Furthermore, a combination of the skewness and the excess kurtosis will enable us to test the binary-origin hypothesis of the stochastic background. These analytical expressions yield multiple constraints that are complementary to standard parameter estimation, such as the Markov chain Monte Carlo method, performed by adopting phenomenological models of SMBHB distribution functions.

The paper is organized as follows. After a brief overview of the energy density spectrum of the gravitational wave background from SMBHBs, Sec.~\ref{sec:SGWBandStats} presents analytical formulas for higher-order statistics of the stochastic background and explains their divergence. Section~\ref{sec:Amplitude_Based_Cutoff} demonstrates how the divergence is resolved by the amplitude threshold based on the sensitivity for individual source detection, which sets a lower limit in redshift for defining the stochastic background. We show that all the higher-order statistics beyond the expectation value depend on the mass function of SMBHBs only through a single weighted average, $\langle \mathcal{M}^{10/3} \rangle$ (see below for definition), once this regularization method is adopted. We also describe our proposals for extracting SMBHB mass information and testing the hypothesis that the stochastic background originates from SMBHBs. Finally, Sec.~\ref{sec:discussion} discusses the implications of our findings and outlines directions for future research. In this work, we adopt a flat $\Lambda$-CDM cosmology with $\Omega_m = 0.3$, $\Omega_{\Lambda} = 0.7$, and $H_0 = 70\,\mathrm{km\,s^{-1}\,Mpc^{-1}}$ when necessary.

\section{Stochastic Gravitational wave Background and Statistics\label{sec:SGWBandStats}}
In this section, we briefly summarize the energy density spectrum of the gravitational wave background $\Omega_{\mathrm{GW}}(f)$ generated by the superposition of contributions from SMBHBs \cite{phinney2001}. Furthermore, we derive expressions for higher-order statistics of the energy density spectrum of the gravitational wave background \cite{lamb2024} and demonstrate that their dependence on the mass and the redshift is expressed concisely in terms of appropriate weighted averages.

\subsection{Energy density spectrum of gravitational waves from SMBHBs\label{subsec:Omega_GW}}
The dimensionless energy density spectrum of the gravitational wave background (hereafter, denoted simply as the energy density spectrum) is defined as
\begin{equation}
\Omega_\mathrm{GW}(f)\equiv \frac{1}{\rho_\mathrm{c}} \frac{d\rho_\mathrm{GW}}{d \ln f}, \label{eq:Omega}
\end{equation}
where $\rho_\mathrm{GW}$ is the energy density of gravitational waves, $\rho_\mathrm{c}$ is the critical density of the Universe, and $f$ is the observed frequency. The energy density spectrum can be represented as a superposition of the events emitting gravitational waves \citep{phinney2001},
\begin{align}
\Omega_\mathrm{GW} (f) & = \frac{1}{\rho_\mathrm{c}} \int^\infty_0\int^\infty_0 \frac{d^2n}{dzd\log {\mathcal M}} \notag\\ 
& \times \frac{1}{1+z} \left. \frac{dE_\mathrm{GW}}{d\ln f_\mathrm{r}} \right|_{f_\mathrm{r}= f(1+z)}dz d\log \mathcal{M},\label{drhodf}
\end{align}
where $dE_\mathrm{GW}/d\ln f_\mathrm{r}$ is the energy spectrum of gravitational waves emitted by a single source in its rest frame and $f_\mathrm{r} = f(1+z)$ is the frequency at the emission. In this paper, we always assume that the gravitational sources are inspiraling SMBHBs. Accordingly, $d^2 n/ dz d\log{\mathcal M}$ specifically means the comoving number density of SMBHB mergers per unit redshift $z$ and per logarithmic chirp mass $\log \mathcal{M}$. The factor $1/(1+z)$ accounts for the cosmological redshift of gravitational waves. 

Rather than directly handling the comoving number density of mergers, statistical quantities are easily derived by an equivalent expression in terms of the frequency distribution of inspiraling binaries \cite{phinney2001}. This approach has been widely adopted to calculate the stochastic gravitational wave background from SMBHBs (see, e.g., Refs.~\cite{sesana2004,ravi2012,sato-polito2024}). The comoving number density of mergers $d^2n / dzd\log\mathcal{M}$ is related to the number of binaries emitting gravitational waves in a logarithmic frequency interval $d^3N/dzd\log\mathcal{M}d\ln f_\mathrm{r}$ via the expression
\begin{align}
\frac{d^2n}{dz d\log \mathcal{M}} = \frac{d^3N}{dz d\log\mathcal{M} d\ln{f_\mathrm{r}}} \frac{dz}{dV_\mathrm{c}}\frac{d\ln{f_\mathrm{r}}}{dt_\mathrm{r}} \frac{dt_\mathrm{r}}{dz},
\label{ntoN}
\end{align}
where $t_\mathrm{r}$ is the time in the source rest frame and $V_\mathrm{c}$ is the comoving volume. The geometric factor is given by $(dz/dV_\mathrm{c}) (dt_\mathrm{r}/dz) = (1+z)/[4\pi c d^2_\mathrm{L}(z)]$, where $d_\mathrm{L}$ is the luminosity distance. In this paper, we assume that binary orbits are quasicircular and that the binaries lose energy solely through gravitational wave emission. In our analysis of higher-order statistics, we focus on the frequency range of $\gtrsim 10\,\text{nHz}$, where the shot noise component becomes prominent (see Fig.~\ref{fig:fluctuation}). Quasicircularity will be appropriate for this high-frequency range because of gravitational radiation reaction \cite{peters1964}, while the appropriate description of the low-frequency regime may require extension to eccentric binaries \cite{enoki2007}. Under these assumptions, $dE_\mathrm{GW}/d\ln f_\mathrm{r}$ and $d \ln f_\mathrm{r}/d t_\mathrm{r}$ are expressed as
\begin{align}
    \frac{dE}{d\ln f_\mathrm{r}} &= \frac{\pi^{2/3}}{3G} (G\mathcal{M})^{5/3} f_\mathrm{r}^{2/3} , \\
    \frac{d\ln f_\mathrm{r}}{dt_\mathrm{r}}&= \frac{96\pi}{5} \left( \frac{\pi G\mathcal{M} f_\mathrm{r}}{c^3} \right)^{5/3} f_\mathrm{r}. \label{dfdt}
\end{align}
We may also define the effective gravitational wave amplitude by
\begin{equation}
     h^2(z,\mathcal{M},f) =\frac{32 \pi^{{4}/{3}}}{5c^8}\frac{(1+z)^{10/3}}{d_\mathrm{L}^2(z)}(G\mathcal{M})^{{10}/{3}}f^{{4}/{3}} . \label{h2}
\end{equation}
Equation \eqref{eq:Omega} can be rewritten by using Eqs.~\eqref{ntoN}, \eqref{dfdt}, and \eqref{h2} to
\begin{align}
    \Omega_\mathrm{GW}(f) = \frac{2\pi^2}{3H_0^2} f^2&\int \frac{d^3N}{dzd\log \mathcal{M} d\ln f_\mathrm{r}} \notag\\
    &\times h^2(z,\mathcal{M}, f) dz d\log \mathcal{M} .\label{omega_N}
\end{align}

\subsection{Higher-order statistics and dependence decomposition \label{subsec:Derivation_of_HOS}}
In this paper, we assume that the comoving number density of SMBHB mergers is separable with respect to chirp mass $\phi(\log \mathcal{M})$ and redshift $\mathcal{E} (z)$,
\begin{align}
    \frac{d^2n}{dzd\log \mathcal{M}}
    &\equiv \dot{n}_0 \phi (\log \mathcal{M}) \mathcal E(z).\label{dndzdM}
\end{align}
The integral of $d^2n/dz d\log\mathcal{M}$ gives the total comoving number density of mergers $n$, i.e.,
\begin{align}
    n&\equiv\int \frac{d^2n}{dz d\log\mathcal{M}} dzd\log\mathcal{M}\notag\\
    &=\dot n_0 \left(\int \phi(\log \mathcal{M}) d\log\mathcal{M}\right) \left(\int \mathcal E(z) dz\right).
\end{align}
In a similar manner to this separated expression, the expectation value and higher-order statistics of the energy density spectrum factorize into weighted averages of chirp-mass and redshift components \cite{phinney2001}. Specifically, we define the weighted average by 
\begin{equation}
    \langle \mathcal{M}^a \rangle 
    \equiv \frac{\int \mathcal M^a\phi(\log \mathcal{M})d \log \mathcal{M}}{\int \phi(\log \mathcal{M})d\log \mathcal{M}} \label{eq:weightedAve_M}
\end{equation}
and
\begin{equation}
     \langle (1+z)^a d_\mathrm{L}^b \rangle \equiv\frac{\int(1+z)^a [d_\mathrm{L} (z)]^b \mathcal{E}(z)dz}{\int \mathcal{E} (z)dz}.\label{eq:weightedAve_z}
\end{equation}

The expectation value of the energy density spectrum $\mathrm{E}[\Omega_\mathrm{GW}(f)]$ corresponds to Eq.~\eqref{drhodf} and, equivalently, Eq.~\eqref{omega_N}. Using Eqs.~\eqref{eq:weightedAve_M} and \eqref{eq:weightedAve_z}, we obtain
\begin{align}
   & \mathrm{E} [\Omega_\mathrm{GW}(f)] \notag \\ 
   &= A_1 f^{{2}/{3}} \int\dot{n}_0 \,\phi(\log \mathcal{M})\, \mathcal{E}(z) \mathcal{M}^{{5}/{3}} (1+z)^{-{1}/{3}} d\log\mathcal{M} dz \notag \\ 
   &= A_1\,f^{2/3}\,n \,\langle\mathcal{M}^{5/3}\rangle\,\langle(1+z)^{-{1/3}}\rangle ,\label{MEANweight}
\end{align}
where $A_1=8 (\pi G)^{5/3}/[9c^2 H_0^2]$ (see also below for $A_{{j}}$). This expression indicates a well-known result that the expectation value is determined by $n$, $\langle \mathcal{M}^{5/3} \rangle$, and $\langle (1+z)^{-1/3} \rangle$ \cite{phinney2001}. Thus, these three cannot be separated only from the averaged spectrum, while the value of $\langle (1+z)^{-1/3} \rangle$ usually depends only weakly on the redshift function.

Formally, higher-order statistics such as the variance, skewness, and excess kurtosis of the energy density spectrum can also be factorized into chirp-mass and redshift components \cite{jaffe2003}. Here, we consider the energy density integrated in the frequency bin with the width $\Delta f$ centered on the frequency $f$, namely, from $f-\Delta f/2$ to $f+\Delta f/2$. Under the condition that the frequency bin is so narrow that the gravitational wave amplitude from a source with given $z$ and $\log\mathcal{M}$ is approximately constant within the interval, the energy density in the frequency bin can be approximated as
\begin{align}
    &\Omega_\mathrm{GW}(f)\Delta \ln f \notag \\
    &\approx \frac{2\pi^2}{3H_0^2}f^2 \sum_{i} \frac{d^3N}{d\theta_{i}\,d\ln f} \,\,h^2(z,\mathcal{M} ,f)\,\,\Delta \theta_{i}\Delta \ln f ,\label{OmegaDeltaf}
\end{align}
where $i$ indicates a hypothetical pixel in a plane spanned by $\theta = \{z , \log \mathcal{M}\}$. Higher-order statistics are conveniently expressed in terms of the cumulant $\kappa_{j}$ defined by
\begin{align}
    \kappa_{j}&[\Omega_\mathrm{GW} (f) ] = (\Delta \ln f )^{1-{j}}\left( \frac{2\pi^2}{3H_0^2} f^2 \right)^{{j}}\notag \\
    &\times \int \frac{d^3N}{dzd\log \mathcal{M} d\ln f_\mathrm{r}} h^{2{j}}(z,\mathcal{M} ,f ) dz d\log \mathcal{M}.
\end{align}

By definition, the first and second cumulants correspond to the expectation value and variance, respectively. 
The third and fourth cumulants, $\kappa_3$ and $\kappa_4$, are related to the skewness and excess kurtosis, respectively (see Appendix~\ref{app:HOS} for the detailed derivation).
By substituting Eqs.~\eqref{ntoN} and \eqref{h2} and employing the expressions for the weighted averages, Eqs.~\eqref{eq:weightedAve_M} and \eqref{eq:weightedAve_z}, the cumulants can be represented by
\begin{widetext}
\begin{align}
    \kappa_{j}[\Omega_\mathrm{GW}(f)] 
    &= A_{j}(\Delta \ln f)^{1-{j}} f^{{(10{j}-8)/3}} \dot{n}_0 
     \int \phi(\mathcal{M}) \mathcal{M}^{{5}(2{j}-1)/3}d\log \mathcal{M} \int \mathcal{E}(z) (1+z)^{(10{j}-11)/3} d_\mathrm{L} ^{-2({j}-1)}dz\label{eq: higher-order}\\
     &= A_{j}(\Delta \ln f)^{1-{j}} f^{{(10{j}-8)}/{3}}  n \langle \mathcal{M}^{{5}(2{j}-1)/3}\rangle \langle (1+z)^{(10{j}-11)/3} d_\mathrm{L} ^{-2({j}-1)} \rangle, \label{eq:kappan_weighted}
\end{align}
\end{widetext}
where the prefactor is given by
\begin{equation}
A_{j}  = \frac{5}{24} \left (\frac{64}{15}\right)^{j} \frac{(\pi G )^{5(2{j}-1)/3}}{H_0^2 c^{8{j}-6}}.
\end{equation}
The first moment, $\kappa_1$, is exactly the same as Eq.~\eqref{omega_N}, but here interpreted as the statistical quantity, the expectation value $\mathrm{E}[\Omega_\mathrm{GW}(f)]$. While the second cumulant, $\kappa_2$, directly corresponds to the variance $\mathrm{V}[\Omega_\mathrm{GW}(f)]$ as stated above, the skewness and excess kurtosis are expressed by combining the cumulants for ${j} \geq 2$. Specifically,
\begin{widetext}
\begin{align}
    \mathrm{V}[\Omega_\mathrm{GW}(f)] &\equiv\kappa_2[\Omega_\mathrm{GW}(f)] 
    = {A_2}\,\frac{f^4}{\Delta{\ln f}}n\,\langle\mathcal{M}^{5}\rangle\,\langle(1+z)^{3}/d^2_\mathrm{L} (z)\rangle, \label{varweight} \\
    \mathrm{S}[\Omega_\mathrm{GW}(f)] &\equiv\frac{\kappa_3[\Omega_\mathrm{GW}(f)]}{\kappa_2[\Omega_\mathrm{GW}(f)]^{3/2}} 
    =  {A_3} \frac{f^{22/3}}{(\Delta \ln f)^2 \mathrm{V}[\Omega_\mathrm{GW}(f)]^{3/2}} n\langle\mathcal{M}^{25/3}\rangle\,\langle(1+z)^{19/3}/d^4_\mathrm{L} (z)\rangle ,\label{skewweight} \\
    \mathrm{K}[\Omega_\mathrm{GW}(f)] & \equiv\frac{\kappa_4 [\Omega_\mathrm{GW}(f)]}{\kappa_2[\Omega_\mathrm{GW}(f)]^2} = {A_4} \frac{f^{32/3}}{(\Delta \ln f)^3 \mathrm{V}[\Omega_\mathrm{GW}(f)]^2} n\langle\mathcal{M}^{35/3}\rangle\,\langle(1+z)^{29/3}/d^6_\mathrm{L} (z)\rangle. \label{kurtweight}
\end{align}
\end{widetext}
Here, the excess kurtosis is defined such that it vanishes for a Gaussian distribution. The frequency dependence of these moments reduces to those derived in Ref.~\cite{lamb2024} once $\Delta \ln f$ is changed to $\Delta f/f$.

\subsection{Divergence of higher-order statistics\label{sec:divergence}}

\begin{figure*}[tbp]
    \includegraphics[width=0.9\linewidth]{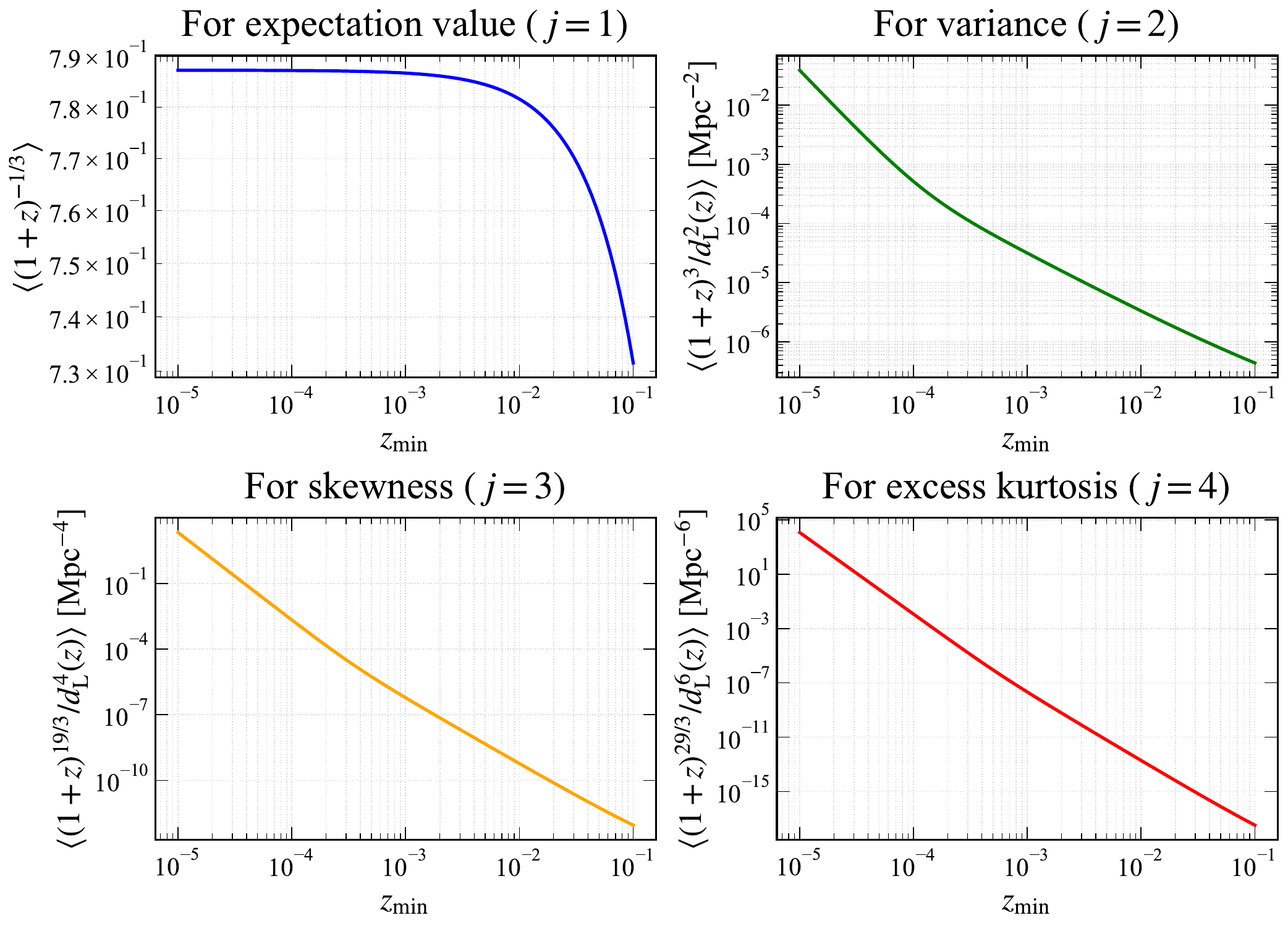}
    \caption{Divergence of higher-order statistics as $z_\mathrm{min} \to 0$. The vertical axis shows the redshift integral of the cumulants defined in Eq.~\eqref{eq:weightedAve_z}, but assumed to be a function of $z_\mathrm{min}$. Except for the expectation value (${j}=1$, left top), the higher-order moments exhibit divergence at $z=0$. The calculation is performed for Model 3 shown in Table~\ref{Modelparam}, and the qualitative behavior is the same for all the models.}
    \label{fig:divergenceofHOS_paper2026}
\end{figure*}

The variance and higher-order statistics of the energy density spectrum are known to diverge if the redshift is integrated down to $z=0$ \cite{jaffe2003}. Figure~\ref{fig:divergenceofHOS_paper2026} shows the redshift components appearing in the expressions for the cumulants $\kappa_{{j}}$ [see Eq.~\eqref{eq:kappan_weighted}], but treated as a function of its lower limit of integration, $z_\mathrm{min}$. While $\kappa_1$ for the expectation value asymptotes to a finite value as $z_{\mathrm {min}}$ approaches zero, all the other cumulants diverge. This divergence is caused by the contribution from individual SMBHBs located in the immediate vicinity of the observer.\footnote{This behavior is not specific to SMBHBs or gravitational waves. Even for a point light source placed at a random position in a Euclidean volume around an observer, higher-order moments of the observed flux diverge, because the probability distribution for the distance $r$ from the observer behaves as $\propto r^2$ and the flux as $\propto r^{-2}$.} The redshift distribution of mergers is frequently modeled by a cutoff power-law function, $\mathcal{E}(z) \propto (1+z)^{\gamma} e^{-(z/z_0)}$ (e.g., \cite{middleton2015,lamb2024,sato-polito2025}) or $\mathcal{E}(z) \propto z^{\gamma} e^{-(z/z_0)^n}$ (e.g., \cite{sato-polito2023,liepold2024,sato-polito2024}). While the resulting integrals converge for sufficiently large $\gamma$ in $\mathcal E(z) \propto z^{\gamma} e^{-(z/z_0)^n}$,
typical values of $\gamma$ adopted in previous studies fall within the regime where the higher-order statistics diverge. For example, models from previous studies such as $\gamma=0.5$ \cite{sato-polito2023} and $\gamma=1.0$ \cite{liepold2024} lead to divergence, rendering an appropriate treatment of the lower limit in the redshift integral essential for any physically meaningful interpretation \cite{lin2026}. 

\begin{figure}[tbp]
    \includegraphics[width=1\linewidth]{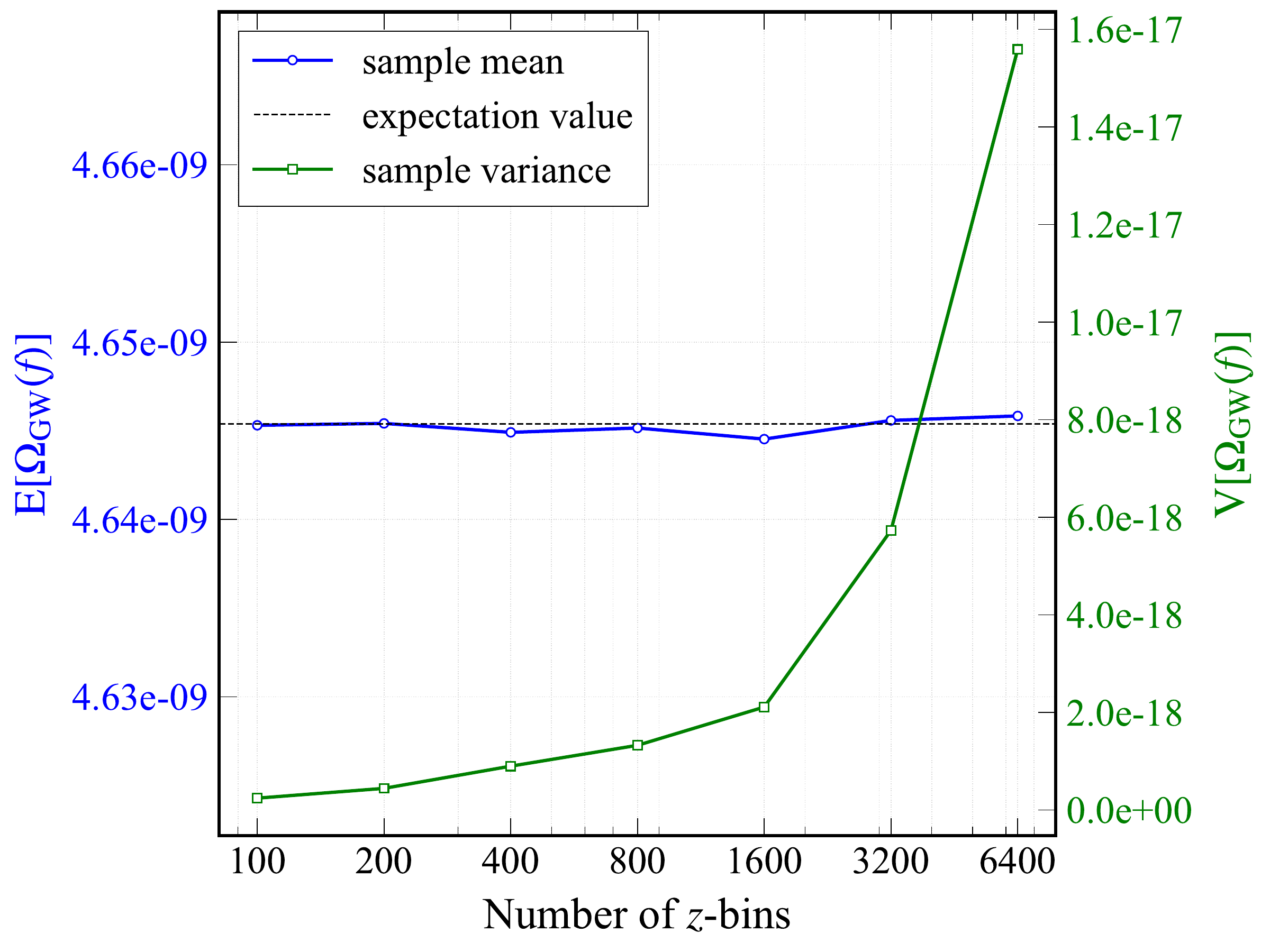}
    \caption{Sample mean (blue, left axis) and sample variance (green, right axis) of the energy density spectrum at $f = 10^{-8}\,\text{Hz}$ and $\Delta \ln f = 0.1$ plotted against the number of redshift bins. The calculation is performed for Model 3 shown in Table~\ref{Modelparam}, and the qualitative behavior is the same for all the models. These numerical results are obtained from $10^7$ Monte Carlo realizations for each number of redshift bins. While the sample mean converges to the expectation value derived from Eq.~\eqref{MEANweight} (dashed line), the sample variance grows without bound. This divergence highlights the necessity of a lower limit in redshift for defining higher-order statistics, which is addressed in the following section.}
    \label{fig:divergenceMeanandVar}
\end{figure}

This divergence also manifests itself as the lack of convergence in Monte Carlo simulations for predicting the gravitational wave background. Typically, these simulations distribute SMBHBs randomly throughout the volume, employing a finite number of redshift bins along with other parameters. This finiteness implicitly imposes a numerical cutoff and yields finite higher-order statistics for a given number of bins. However, these higher-order statistics grow without bound as the number of bins increases, as demonstrated in Fig.~\ref{fig:divergenceMeanandVar}. While the sample mean converges toward a finite, analytically predicted value as the number of redshift bins increases, the sample variance grows without bound. The sample skewness and excess kurtosis similarly grow without bound. This behavior indicates that the resulting higher-order statistics are governed by numerical resolution in the redshift integration rather than physical properties of the source population, making these values meaningless unless a physically motivated regularization is applied. We stress that, as described in this subsection, this problem is not only specific to numerical computations but also applicable to observed quantities.

\section{Regularization of higher-order statistics \label{sec:Amplitude_Based_Cutoff}}
In order to extract physically meaningful information from higher-order statistics of the gravitational wave background, it is essential to establish a method for appropriately defining these statistics. To this end, we need to impose a finite lower limit $z_{\mathrm {min}}$ on the redshift integral. As repeatedly pointed out in the literature, however, an artificial cutoff will result in equally artificial interpretation \cite{ellis2023, ellis2024, sato-polito2024, raidal2026, ali-haimoud2026}. In this section, we propose a physically motivated scheme and its application.

\subsection{Cutoff based on the individual source detection}

In this study, we propose an observationally grounded approach for determining the lower limit in redshift, namely a sensitivity-based threshold derived from the PTA’s detection limit for individual sources (see also Ref.~\cite{ali-haimoud2026}). In the actual data analysis, sufficiently high-amplitude SMBHBs will be resolved as individual (continuous) sources \cite{agazie2023}, akin to the LIGO-Virgo-KAGRA detection of (coalescing) stellar-mass binary black holes. This classification, based on the observational sensitivity for individual sources, serves as the natural and perhaps necessary definition of the genuine gravitational wave background constituted by the ensemble of unresolved sources. Thus, binaries sufficiently close to $z=0$ will always be excluded from the gravitational wave background. This regularization can be imposed solely by gravitational wave data analysis, providing a self-consistent, nonartificial framework for the statistical characterization of the background population.

\begin{figure}[tbp]
    \centering
    \includegraphics[width=1\linewidth]{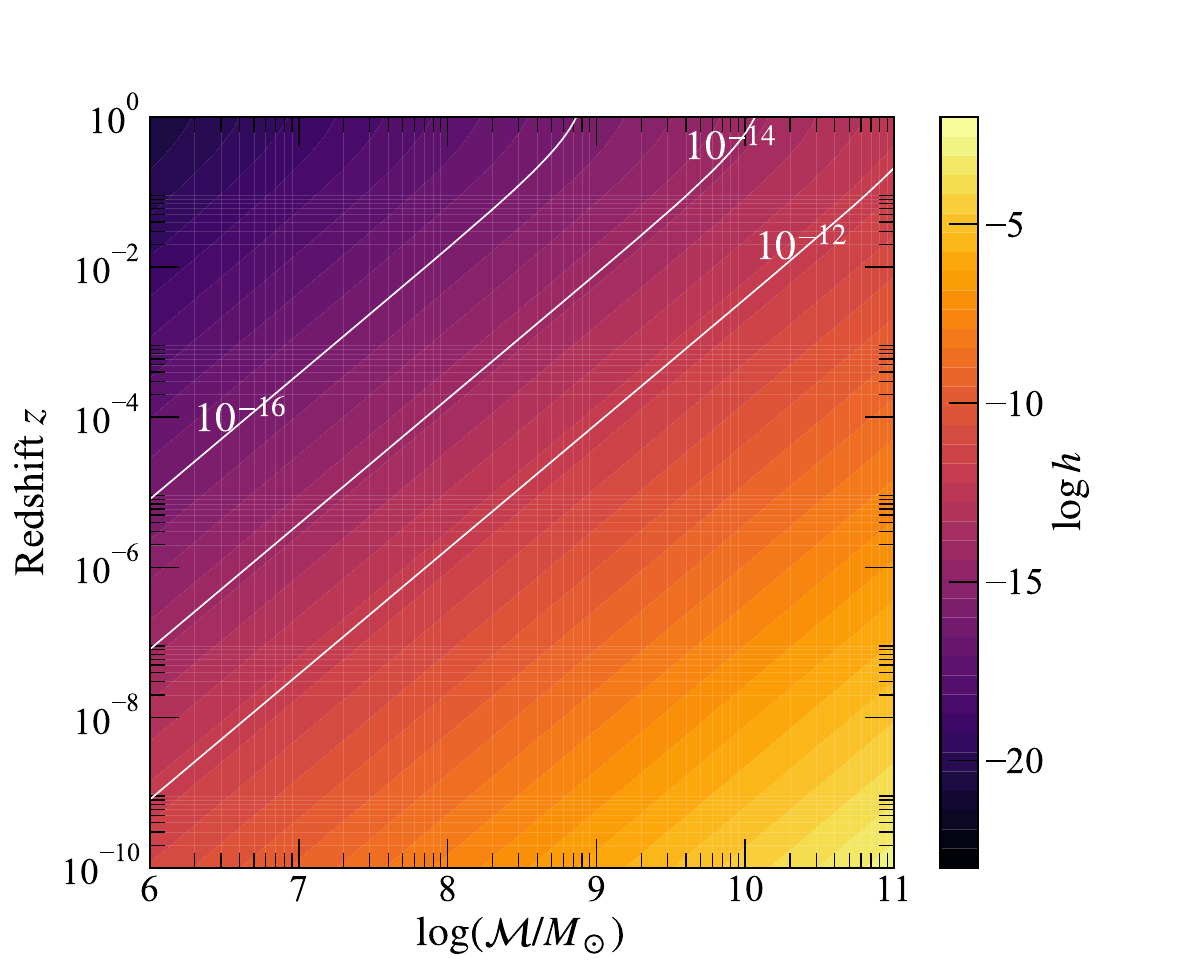}
    \caption{Contour of effective gravitational wave amplitude defined by Eq.~\eqref{h2}. The results are shown for $f = 10^{-8}\,\text{Hz}$, and $h \propto f^{2/3}$. This plot shows that the lower limit of the redshift integral, when set based on the threshold for the gravitational wave amplitude, is an increasing function of the chirp mass, $\mathcal{M}$.}
    \label{fig:h_th_contour}
\end{figure}

By defining $h_\mathrm{th}(f)$ as the threshold amplitude for individual source detection, the lower limit of the redshift integration becomes a function of both the chirp mass and frequency, denoted by $z_\mathrm{min}(\mathcal{M},f)$. Here, we neglect dependence on the orbital inclination, assuming that it is weak \cite{ali-haimoud2026}. Figure~\ref{fig:h_th_contour} shows the effective gravitational wave amplitude $h$ defined by Eq.~\eqref{h2} as a function of the chirp mass $\mathcal{M}$ and the redshift $z$. This figure also illustrates the relation between the detection threshold $h_\mathrm{th}$ and the lower limit $z_\mathrm{min}$, where $z_\mathrm{min}$ increases with the chirp mass for a fixed value of $h_\mathrm{th}$. This mass-dependent cutoff effectively regularizes the statistical moments by excluding resolvable nearby sources. Because all the cumulants and higher-order statistics are derived for each frequency, we suppress the frequency dependence of $h_{\mathrm {th}}(f)$ in the following for ease of notation.

\begin{figure*}[tbp]
    \includegraphics[width=0.9\linewidth]{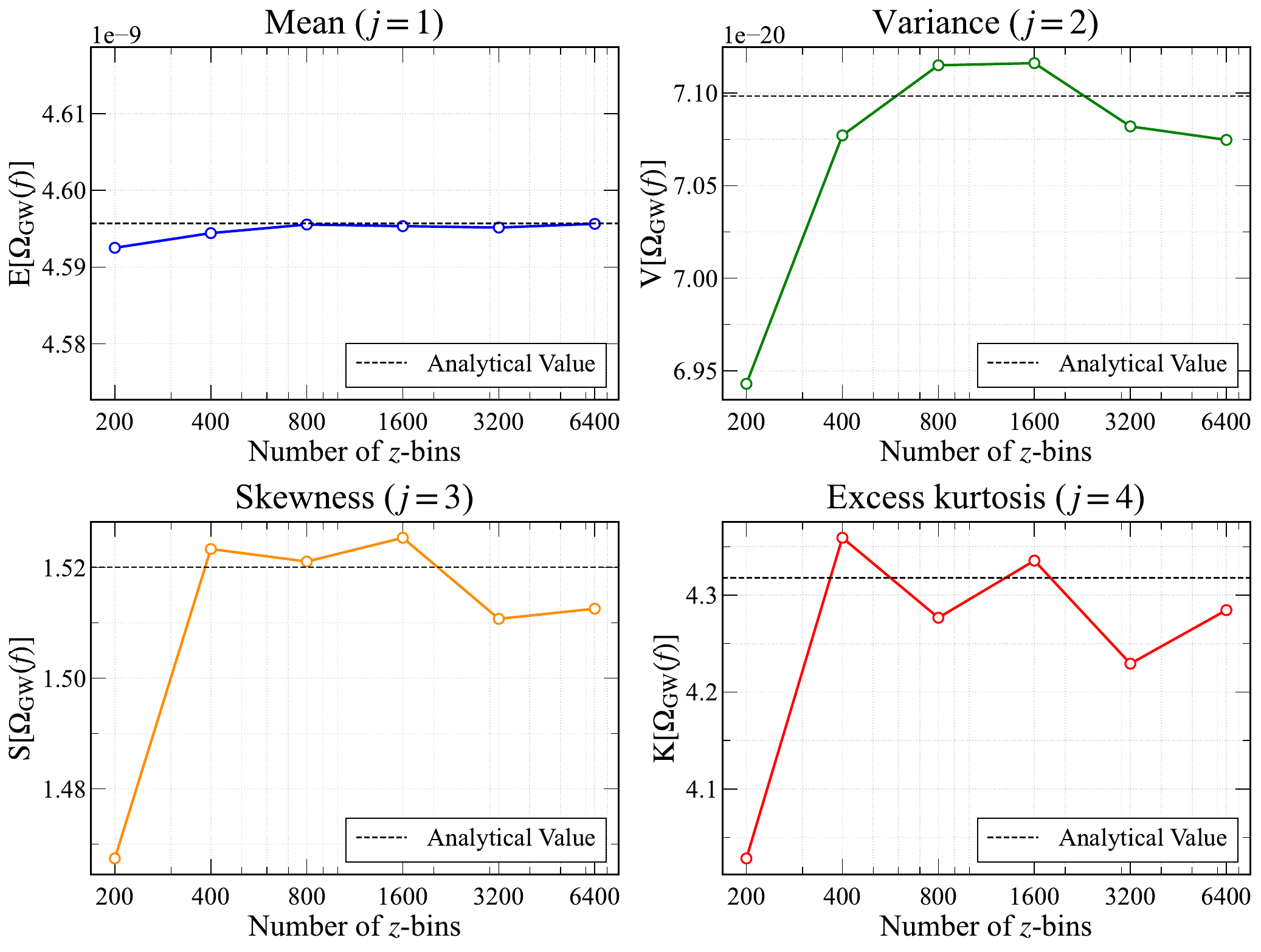}
    \caption{Convergence of higher-order statistics at $f= 10^{-8} \, \text{Hz}$ and $\Delta \ln f = 0.1$ with the amplitude threshold $h_\mathrm{th} = 10^{-15}$. The data points represent higher-order moments obtained from $10^7$ Monte Carlo realizations with different numbers of redshift bins in the same manner as those performed without imposing the threshold for Fig. \ref{fig:divergenceMeanandVar}. The dashed line shows the analytical value obtained by Eq.~\eqref{eq: higher-order} but imposing $z_\mathrm{min}$. Unlike the divergent case in Fig. \ref{fig:divergenceMeanandVar}, higher-order moments, and thus, the cumulants $\kappa_{j}\, ({j}\ge2)$, appear to converge as the number of redshift $z$ bins increases.}
    \label{fig:amplitudebasedconvergence}
\end{figure*}

After establishing a lower limit based on individual source detection, we confirm in Fig.~\ref{fig:amplitudebasedconvergence} that the higher-order statistics become convergent in Monte Carlo simulations. We adopt $h_\mathrm{th} = 10^{-15}$ and $f= 10^{-8}\, \text{Hz}$ to exclude sources with $h>h_\mathrm{th}$. In contrast to the behavior observed in Fig.~\ref{fig:divergenceMeanandVar}, all the statistical moments, including higher ones, converge as the number of redshift bins increases.

\subsection{Lowest-order approximation\label{subsec:h_approx}}

Our primary motivation for considering higher-order statistics is to extract information about the SMBHB population independently of specific models for the distribution function. To gain insight into extractable information, it is useful to adopt a low-redshift approximation $1+z \approx 1$ and $d_\mathrm{L}(z) \approx cz/H_0$. This is reasonably justified by the facts that (i) individually resolvable sources are typically located in the relatively nearby Universe so that $z_\mathrm{min} (\mathcal{M},f) \ll 1$ for a wide range of parameters, and (ii) the reason for the divergence of higher-order statistics at $z=0$ is the dominant contribution from sources at $z \ll 1$. In this approximation, the threshold condition $h(\mathcal{M},f) =h_\mathrm{th}$ with Eq.~\eqref{h2} yields
\begin{equation}
      z_\mathrm{min}(\mathcal{M},f)
      \approx \frac{H_0}{c}\left( \frac{32 \pi^{{4/3}}}{5c^8}\frac{1}{h^2_\mathrm{th}}G^{10/3}f^{4/3}\right)^{{1/2}}\mathcal{M}^{{5/3}} .\label{z(M)min}
\end{equation}
This allows us to approximate the redshift integral by
\begin{align}
    &\int _{z_\mathrm{min}(\mathcal{M},f)}^{z_\mathrm{max}} \mathcal{E}(z) (1+z)^{\frac{1}{3}(10{j}-11)} d_\mathrm{L}^{-2({j}-1)}dz \notag \\
    &\approx \left. \mathcal{E} (z) \right|_{z\to 0} \left( \frac{c}{H_0}\right)^{-2({j}-1)} \frac{1}{2{j}-3} z_\mathrm{min}(\mathcal{M},f)^{-2{j}+3} \notag \\
    &\propto \left( f^{{2}/{3}} \mathcal{M}^{{5}/{3}}\right)^{-2{j}+3},\quad({j}\ge2).
\end{align}
Substituting this result into the general expression for the cumulants in Eq.~\eqref{eq: higher-order} and integrating over the mass function, we obtain
\begin{widetext}
\begin{align}
    & \kappa_{j}[\Omega_\mathrm{GW}(f)] \notag \\
    &{\approx} A_{j} \, \mathcal{E}(z) |_{z\to 0} \frac{1}{2{j}-3} \frac{H_0}{c}\  
    \left (\frac{32 \pi^{4/3}}{5c^8}\frac{1}{h^2_\mathrm{th}}G^{10/3} \right )^{(-2{j}+3)/2} (\Delta \ln f)^{1-{j}} f^{(6{j}-2)/3}\,\dot{n}_0\int d\log \mathcal{M} \phi(\mathcal{M}) \mathcal{M}^{10/3} \label{eq:kappa_amplitude}\\
    &= A_{j} \, \frac{\mathcal{E}(z)|_{z\to 0}}{\int \mathcal{E}(z) dz}  \frac{1}{2{j}-3} \frac{H_0}{c}\
   \left(\frac{32 \pi^{4/3}}{5c^8}\frac{1}{h^2_\mathrm{th}}G^{10/3} \right)^{(-2{j}+3)/2}   (\Delta \ln f)^{1-{j}} f^{(6{j}-2)/3}\,n\,\langle \mathcal{M}^{10/3} \rangle,\quad ({j}\ge2) .\label{eq:kappa_amplitude_weighted}
\end{align}
\end{widetext}
It is remarkable that, at the lowest-order approximation, all the higher-order cumulants with ${j}\ge2$ depend on the mass function $\phi (\log \mathcal{M})$ only through $\langle \mathcal{M}^{10/3}\rangle$, irrespective of the specific functional form. In light of the fact that $\langle \mathcal{M}^{10/3} \rangle$ also governs the probability distribution function at an infinitely large number of sources \cite{ali-haimoud2026}, which is complementary to the current discussion of statistical fluctuations caused by shot noise, we may safely say that $\langle \mathcal{M}^{5/3} \rangle$ for the expectation value and $\langle \mathcal{M}^{10/3} \rangle$ play distinctive roles in determining the gravitational wave background.

We demonstrate that $\langle \mathcal{M}^{10/3} \rangle / \langle \mathcal{M}^{5/3} \rangle$ can be inferred with moderate precision by combining the mean and variance (or any higher-order moments) of the observed gravitational wave background. We recall that the expectation value, i.e., $\kappa_1$, does not suffer from divergence. Therefore, we continue to adopt the expectation value of the full SMBHB population defined by integrating down to $z=0$. This choice is convenient because the mass and redshift integrations remain separable even without adopting the lowest-order approximation, allowing us to express $\kappa_1$ as $\kappa_1 = A_1 f^{2/3} n \langle \mathcal{M}^{5/3} \rangle \langle (1+z)^{-1/3} \rangle$. In the actual data analysis, this corresponds to counting both the background and the individual sources for estimating the total energy density of gravitational waves. We may equally adopt the expectation values obtained by integrating down to $z=z_\mathrm {min}$. Both definitions are formally equivalent in the lowest-order approximation adopted here. In Sec.~\ref{subsec:result_h}, we discuss quantitatively how the expectation value depends on the choice of these integration limits by relaxing the lowest-order approximation for some specific models. 

By taking the ratio of the higher-order cumulants $\kappa_{j}$ (for ${j}\ge 2$) under the lowest-order approximation from Eq.~\eqref{eq:kappa_amplitude_weighted} to $\kappa_1$, we obtain 
\begin{align}
    \frac{\kappa_{j}}{\kappa_1} &=\frac{A_{j}}{A_1} \frac{1}{2{j}-3} \frac{H_0}{c} \left( \frac{32 \pi^{4/3}}{5c^8}\frac{1}{h^2_\mathrm{th}}G^{10/3} \right)^{(-2{j}+3)/2} \notag \\
    \times & \frac{{\mathcal{E}(z)}|_{z\to 0}}{\int (1+z)^{-1/3} \mathcal{E}(z)dz} (\Delta \ln f)^{1-{j}} f^{(6{j}-4)/3}  \frac{\langle \mathcal{M}^{10/3} \rangle}{\langle \mathcal{M}^{5/3} \rangle}\\
    &\propto \frac{{\mathcal{E}(z)}|_{z\to 0}}{\int (1+z)^{-1/3} \mathcal{E}(z)dz} \frac{\langle \mathcal{M}^{10/3} \rangle}{\langle \mathcal{M}^{5/3} \rangle} .\label{MeanVarRatio}
\end{align}
Notably, the total comoving number density of mergers, $n$, and thus $\dot{n}_0$ cancels out in this ratio. This should be contrasted with $\kappa_1$, which cannot resolve the degeneracy between $\dot{n}_0$ and $\langle \mathcal{M}^{5/3} \rangle$ by its own. This demonstrates that higher-order statistics could serve as a direct probe of the mass function $\phi(\log \mathcal{M})$, independent of the absolute normalization of the merger rate. The drawback is that this ratio depends on the redshift function via $\mathcal{E}(0) / \int (1+z)^{-1/3} \mathcal{E}(z) dz$. To make matters worse, this factor depends more sensitively on the redshift function $\mathcal{E}(z)$ than $\langle (1+z)^{-1/3} \rangle$ characterizing the expectation value. We return to this issue in the next subsection by assuming specific distribution function models.

While the same mass scaling, $\langle \mathcal M^{10/3} \rangle$, in the lowest-order approximation might appear to suggest that the skewness and excess kurtosis provide no additional information beyond the variance, this is not necessarily the case. First, by relaxing the lowest-order approximation and accounting for the full redshift evolution $\mathcal E(z)$ as well as the exact $d_\mathrm{L}(z)$, the skewness and excess kurtosis acquire dependence on the mass and redshift functions in a manner different from the variance. The extent to which we may exploit the full dependence on $z$ depends on the observational errors, and we leave this investigation for our future study.

Second, even at the lowest-order approximation, by combining the cumulants so that the dependence on the distribution function cancels, they can be utilized as a diagnostic tool for testing the hypothesis that the gravitational wave background originates from SMBHBs. As a specific example, we take the ratio of the excess kurtosis to the squared skewness, $\mathrm{K}/\mathrm{S}^2$, which is given by a combination of three cumulants, $\kappa_2 \kappa_4 / \kappa_3^2$. Equation~\eqref{eq:kappa_amplitude_weighted} readily shows that all the physical quantities including the distribution function completely cancel in this expression, yielding a model-independent consistency relation,
\begin{align}
     \frac{\mathrm{K}}{\mathrm{S}^2} = \frac{\kappa_2 \kappa_4}{\kappa_3^2} =\frac{9}{5} , \label{eq:kpers2}
\end{align}
even irrespective of the frequency. Indeed, inequalities often hold between fourth moments like excess kurtosis and squared third moments like skewness, such as Pearson's inequality in statistics and Suyama-Yamaguchi inequality for non-Gaussianity in inflationary cosmology \cite{2008PhRvD..77b3505S}. Equation \eqref{eq:kpers2} indicates that this gravitational wave background obeys a leptokurtic distribution, and thus, obviously satisfies Pearson's inequality, $\mathrm{K} \ge \mathrm{S}^2 - 2$. Our consistency relation is an equality derived under the lowest-order approximation, akin to Suyama-Yamaguchi inequality saturated for a single degree of freedom.

If future PTA observations demonstrate that this equality shows a large deviation from 1.8, i.e., by a factor of a few, it will strongly disfavor the hypothesis that the gravitational wave background originates from SMBHBs.
While environmental effects or orbital eccentricity might cause deviations in the low-frequency range, they may not be very effective in the high-frequency range, where higher-order statistics could be detected with significance. Although the value $9/5$ changes once the lowest-order approximation is relaxed, the variation is likely to be much smaller than conceivable observational errors. In the next subsection, we verify this statement by adopting specific models of the distribution function.

\subsection{Application to specific models\label{subsec:result_h}}

\begin{table*}[tbp]
\caption{Parameters of the SMBHB distribution models. The functional form is given by Eq.~\eqref{presssche}. The values of $(\alpha, \mathcal{M}_*, \beta, z_0)$ are adopted from Ref.~\cite{sato-polito2025}, and those of $\dot{n}_0$ from Ref.~\cite{lamb2024}. We also show the values for $\langle \mathcal{M}^{5/3} \rangle^{3/5}$, $\langle \mathcal{M}^{10/3} \rangle^{3/10}$, $\langle (1+z)^{-1/3} \rangle$, and $\mathcal{E}(0) / \int \mathcal{E}(z) (1+z)^{-1/3} dz$.}
\label{Modelparam}
\centering
\renewcommand{\arraystretch}{1.5}
\begin{tabular}{c|ccccc|cccc}
\hline
& $\dot n_0[\mathrm {Mpc^{-3} Gyr^{-1}}]$& $\alpha$ & $\mathcal M_*[M_\odot]$ & $\beta$ & $z_0$ & $\langle \mathcal{M}^{5/3} \rangle^{3/5} [M_\odot]$ & $\langle \mathcal{M}^{10/3} \rangle^{3/10} [M_\odot]$ & $\langle (1+z)^{-1/3} \rangle$ & $\mathcal{E}(0) / \int \mathcal{E}(z)(1+z)^{-1/3} dz$ \\
\hline \hline
Model 1 & $2.0\times 10^{-2}$ & $1$ & $3.2\times10^7$ & $3$ & $3$ & $4.84 \times 10^6$ & $1.24 \times 10^7$ & $0.702$ & $0.215$ \\
\hline
Model 2 & $9.5\times 10^{-3}$ & $0.5$ & $7.5\times10^7$ & $2.5$ & $2.4$ & $1.47 \times 10^7$ & $4.00 \times 10^7$ & $0.741$ & $0.403$ \\
\hline
Model 3 & $1.5\times 10^{-3}$ & $0$ & $1.8\times10^8$ & $2$ & $1.8$ & $6.76 \times 10^7$ & $1.55 \times 10^8$ & $0.787$ & $0.721$ \\
\hline
Model 4 & $9.0\times 10^{-5}$ & $-0.5$ & $4.2\times 10^8$ & $1.5$ & $1.1$ & $3.23 \times 10^8$ & $5.79 \times 10^8$ & $0.846$ & $1.34$ \\
\hline
Model 5 & $2.0\times 10^{-6}$ & $-1$ & $1.0\times 10^9$ & $1$ & $0.5$ & $1.28 \times 10^9$ & $1.95 \times 10^9$ & $0.908$ & $2.79$ \\
\hline
\end{tabular}
\end{table*}

To make the discussion quantitative, we adopt a specific model for the SMBHB distribution (see, e.g., \cite{middleton2015,sato-polito2025})
\begin{equation}
    \frac{d^2n}{dzd\log \mathcal{M}} = \dot{n}_0\left(\frac{\mathcal{M}}{10^7 M_\odot} \right)^{-\alpha} e^{-\mathcal{M}/\mathcal{M}_*} (1+z)^{\beta} e^{-{z/z_0}}\frac{dt_\mathrm{r}}{dz},\label{presssche}
\end{equation}
where $dt_\mathrm{r}/dz = [(1+z) H(z)]^{-1}$. For the purpose of investigating various possibilities, we consider several parameter sets that represent a broad range of SMBHB distributions presented in Table~\ref{Modelparam}. This model is described by five parameters. The parameter $\dot n_0$ is the normalized merger rate. The parameter $\alpha$ expresses the low-mass power law and the parameter $\mathcal M_*$ expresses the high-mass cut off for the distribution. This mass function is known as the Schechter function. The parameters $\beta$ and $z_0$ provide a similar description for the redshift distribution. This model is often chosen for estimating the expectation value of the merger rate of SMBHBs without assuming a particular merger history. We consider binaries in the chirp-mass and redshift ranges $10^6 M_\odot \leq \mathcal M \leq 10^{11}M_\odot$ and $0\leq z \leq 5,$ respectively. For the higher-order cumulants, however, the effective lower limit of the redshift integration is $z_{\mathrm{min}}(\mathcal M,f ,h_{\mathrm {th}})>0$, as described in Sec.~\ref{subsec:h_approx}.  Model 1 represents a large population of low-mass binaries, and Model 5 represents a sparse population of high-mass binaries. We also present various relevant weighted averages as well as $\mathcal{E}(0) / \int \mathcal{E}(z) (1+z)^{-1/3} dz$. We comment on a possible alternative model later.

As discussed in the previous subsection, Eq.~\eqref{MeanVarRatio} retains dependence on the redshift function via $\mathcal{E}(0)/\int \mathcal{E}(z)(1+z)^{-1/3} dz$. The last column of Table~\ref{Modelparam} shows that this term varies by a factor of $\approx 15$ among the models adopted here. If we scale Eq.~\eqref{MeanVarRatio} to give it a mass dimension, this uncertainty translates to a factor of $\approx 5$. While this systematic uncertainty will not be critical as long as the statistical uncertainty of variance estimation is $\gtrsim 15$, it may become problematic as observational precision improves. It should also be commented that, while $\langle (1+z)^{-1/3} \rangle$ varies by only $\approx 30\%$, the uncertainty in $n$ (not shown here) exceeds 4 orders of magnitude. The latter renders clear extraction of $\langle \mathcal{M}^{5/3} \rangle$ from the observed mean unrealistic.

\begin{figure*}[tbp]
    \centering
    \includegraphics[width=1\linewidth]{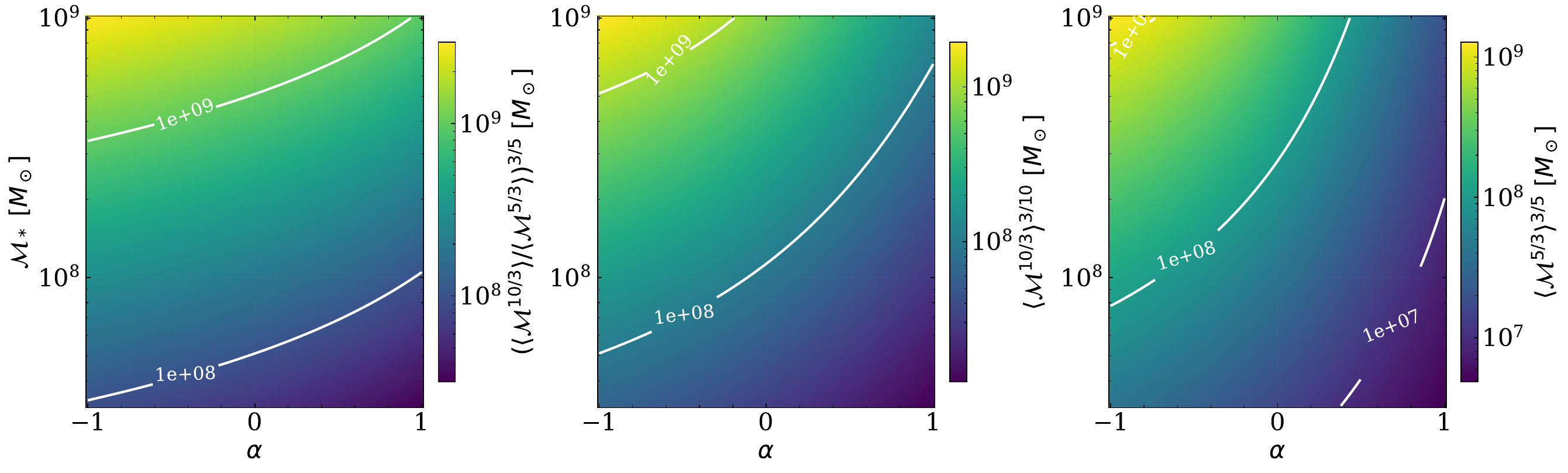}
    \caption{Contour plots of weighted averages of the chirp mass and their ratios for given mass functions, shown in the parameter space of $\alpha$ and $\mathcal M_{*}$. From left to right, the panels represent the contours for $\langle \mathcal{M}^{10/3} \rangle / \langle \mathcal{M}^{5/3} \rangle$, $\langle \mathcal{M}^{10/3} \rangle$, and $\langle \mathcal{M}^{5/3} \rangle$, respectively. These quantities are scaled to have units of $M_\odot$. While $\alpha$ and $\mathcal{M}_{*}$ are degenerate in determining the higher-order statistics (center) and the expectation value (right), the degeneracy is modified to mitigate dependence on $\alpha$ in their ratio (left). The resulting flattened contours demonstrate that $\mathcal{M}_{*}$ can be effectively constrained.}
    \label{fig:contour}
\end{figure*}

If we assume that the dependence on the redshift function can be neglected, the ratio of observed higher-order moments to the observed mean is essentially determined by $\langle \mathcal{M}^{10/3} \rangle / \langle \mathcal{M}^{5/3} \rangle$. Figure \ref{fig:contour} illustrates how these mass-weighted averages depend on the model parameters. The left panel shows the ratio $\langle \mathcal{M}^{10/3} \rangle / \langle \mathcal{M}^{5/3} \rangle$, and the middle and right panels plot the numerator and the denominator, respectively. Because each plot shows the contour for a single variable, it is evident that the parameters for the mass distribution, $\alpha$ and $\mathcal{M}_*$, are degenerate. However, by taking the ratio of the higher-order statistics to the mean, the nature of this degeneracy changes, and the contours flatten out. This demonstrates that taking the ratio significantly enhances the sensitivity to $\mathcal{M}_*$ characterizing the upper end of the mass distribution. Dependence on the redshift function will likely introduce uncertainties comparable to those induced by $\alpha$. 

As we commented in Sec.~\ref{subsec:h_approx}, we also compare the expectation value obtained by integrating down to $z=0$ with that obtained after applying the same individual source threshold used for the higher-order cumulants. For $h_{\mathrm {th}} = 10^{-14}$, which is the current PTA's amplitude threshold at $f = 10^{-8}~{\mathrm {Hz}}$, the difference between the two  definitions is $10^{-4}\%$ -- $1\%$ for Models 1-4, while it reaches $13\%$ for Model 5. Thus, the difference may be non-negligible if a large number of high-mass or nearby sources are present. The difference becomes larger for $h_{\mathrm {th}} = 10^{-15}$, reflecting the increasing contribution removed from the expectation value as more nearby sources become individually resolvable. We recall that either definition can be adopted in the actual data analysis.

We also comment on a possible alternative model of the mass function~\cite{sesana2008},

\begin{align}
\phi(\log \mathcal M)=
\begin{cases}
 \left ({\mathcal M}/{10^7 M_\odot} \right )^{-\alpha} \quad& (\mathcal M_{\mathrm {min}} \leq \mathcal M \leq \mathcal M_{\mathrm {max}} )\\
0 \quad &(\mathrm {else})
\end{cases},
\end{align}
which is characterized by three parameters $\alpha, \, \mathcal M_{\mathrm {min}},$ and $\mathcal M_{\mathrm {max}}$. In this case, the ratio $\langle \mathcal{M}^{10/3} \rangle / \langle \mathcal{M}^{5/3} \rangle$ can be derived analytically, and it is approximated for $\mathcal M_{\mathrm {min}}/\mathcal M_{\mathrm {max}} \ll 1$ as

\begin{equation}
\frac{\langle \mathcal{M}^{10/3} \rangle }{ \langle \mathcal{M}^{5/3} \rangle }\sim\left ( 1 + \frac{5}{3\alpha -10}\right )\mathcal{M}_\mathrm{max}^{5/3}.
\end{equation}
This expression indicates that this ratio is determined primarily by $\mathcal M_{\mathrm {max}}$. Thus, even for this alternative mass function, the ratio $\langle \mathcal M^{10/3}\rangle/\langle \mathcal M^{5/3}\rangle$ is primarily sensitive to the high-mass end of the mass function. This demonstrates that the sensitivity of $\kappa_2/ \kappa_1$ to the high-mass end is robust with respect to the assumed form of the mass function.
\begin{figure}
    \includegraphics[width=1\linewidth]{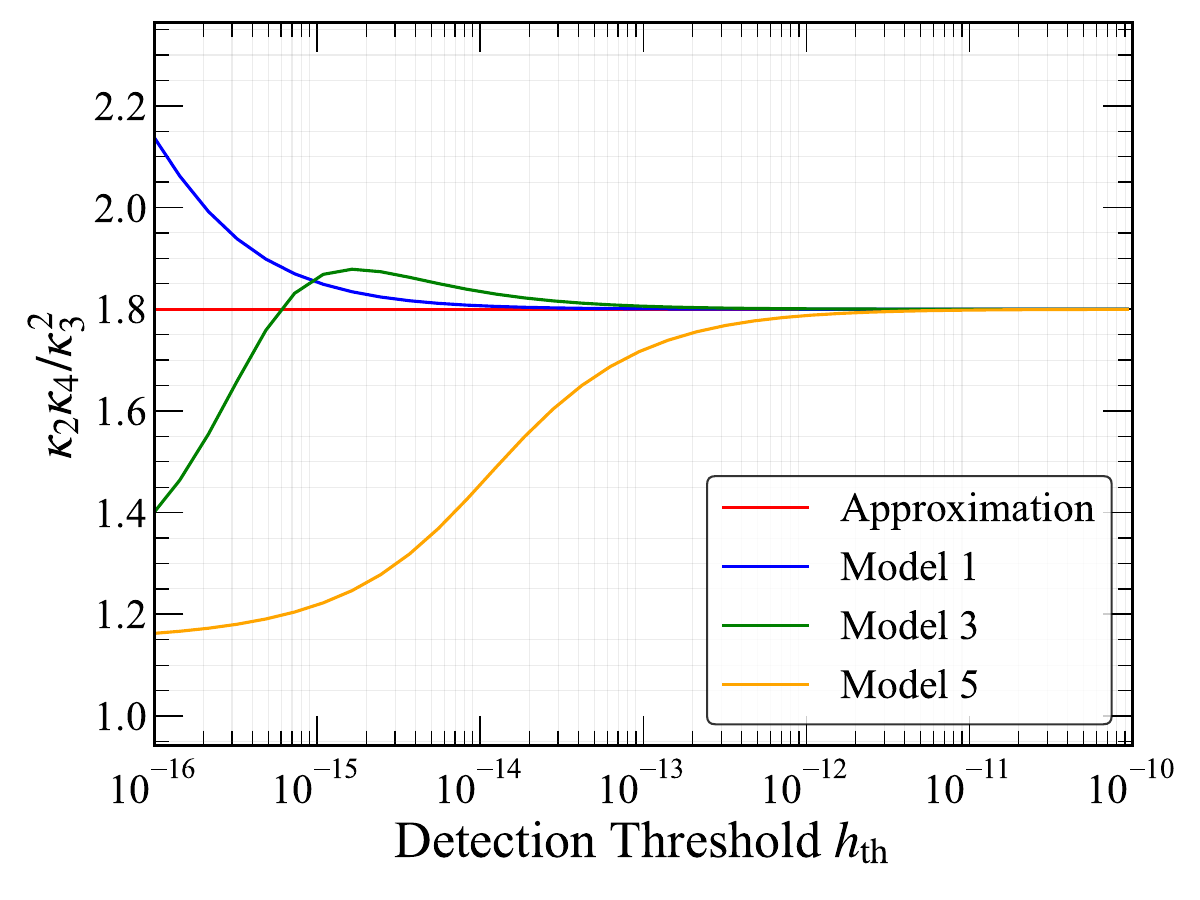}
    \caption{Ratio of cumulants $\kappa_2 \kappa_4 / \kappa_3^2$ plotted against the amplitude threshold $h_\mathrm{th}$. The red solid line represents the value obtained from the lowest-order approximation, $9/5$. The other curves show the exact numerical results for the three models described in Table~\ref{Modelparam}. Even in the exact treatment, the ratio remains close to $9/5$, with a maximum deviation of $\sim 30\%$ for $h_\mathrm{th} \ge 10^{-14}$.}
    \label{fig:kappa_ratio}
\end{figure}

Figure \ref{fig:kappa_ratio} compares the analytical value, $9/5$, for $\kappa_2 \kappa_4 / \kappa_3^2$, or $\mathrm{K}/\mathrm{S}^2$, obtained under the lowest-order approximation with those from exact numerical integration for representative models. Although the exact treatment reveals that cumulants at different orders possess distinct sensitivities to the distribution function, our numerical results show that $\kappa_2 \kappa_4 / \kappa_3^2$ deviates from $9/5$ by, at most, $\sim 30\%$ across all considered models for the current PTA's amplitude threshold $h_\mathrm{th} \gtrsim 10^{-14}$ for detecting individual sources \cite{agazie2023}. This insensitivity implies that {the observed ratio should not deviate by, e.g., a factor of a few, }from $9/5$ if the gravitational wave background originates from SMBHBs.
Consequently, a substantial departure of observed $\mathrm{K}/\mathrm{S}^2$ from this predicted value {would provide grounds for rejecting the binary-origin hypothesis.}
While quantifying the expected ratios for alternative sources such as early-Universe topological defects or secondary gravitational waves remains a subject for future work, the key takeaway is that combinations like $\mathrm{K}/\mathrm{S}^2$ provide a diagnostic tool.

\begin{figure*}
    \centering
    \includegraphics[width=0.9\linewidth]{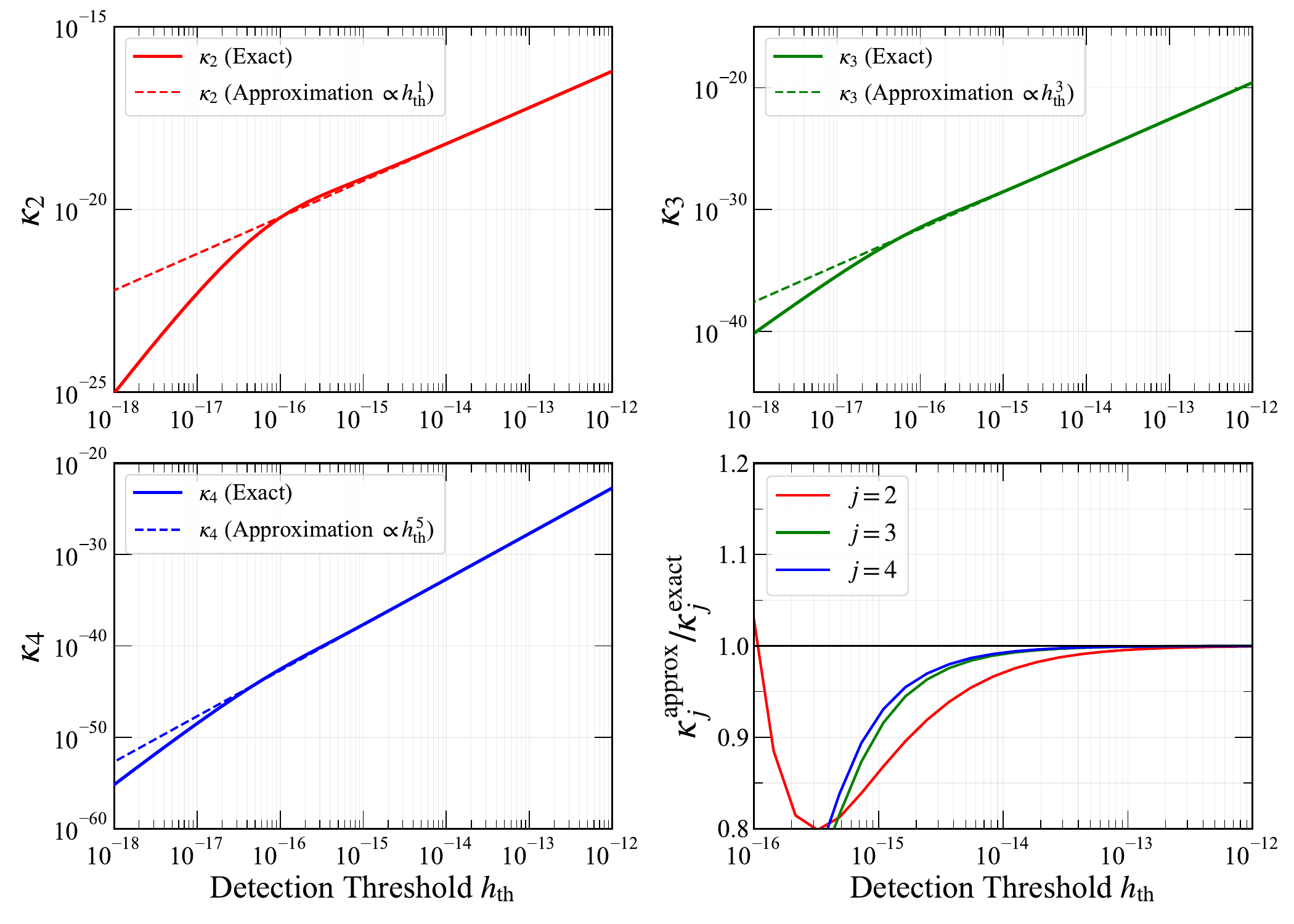}
    \caption{Dependence of $\kappa_{j}$ on the amplitude threshold $h_\mathrm{th}$ for individual source detection. The top-left, top-right, and bottom-left panels show $\kappa_2$, $\kappa_3$, and $\kappa_4$, respectively, for Model 3 and $f=10^{-8}\,\text{Hz}$. In each panel, the exact results (solid curves) are compared with the lowest-order approximations (dashed lines). The bottom-right panel displays the ratio of the approximated values to the exact ones, $\kappa_{j}^\mathrm{approx} / \kappa_{j}^\mathrm{exact}$, for $\kappa_2$, $\kappa_3$, and $\kappa_4$. The error associated with the approximation increases as the threshold $h_\mathrm{th}$ decreases. For the current PTA's sensitivity, $h_\mathrm{th} \gtrsim 10^{-14}$, the relative error is approximately a few percent.}
    \label{fig:kappa_appVSExact}
\end{figure*}

Finally, we briefly assess the validity of the lowest-order approximation. Figure~\ref{fig:kappa_appVSExact} compares the exact numerical integrations with numerical integrations at the lowest-order approximation for a range of the amplitude threshold, $h_\mathrm{th}$, for individual source detection. Here, we adopt Model 3 with intermediate properties and $f = 10^{-8}\,\text{Hz}$. As expected, the higher the amplitude threshold, the better the accuracy of the approximation. The current PTA's sensitivity depends on the frequency, and generally $h_\mathrm{th} \gtrsim 10^{-14}$ \cite{agazie2023}. In this situation, the discrepancy is only a few percent between the exact result and the lowest-order approximation. We also show results for Models 1 and 5 in Appendix~\ref{app:cumulants_and_ratios}.

\section{discussion and conclusion\label{sec:discussion}}
In this paper, we propose a physically meaningful definition of higher-order statistics for the stochastic gravitational wave background by introducing a lower limit $z_\mathrm{min}$ for redshift integration. Instead of putting an artificial cutoff such as $z_\mathrm{min} = \text{const.}$, we adopt a lower limit, $z_\mathrm{min} (\mathcal{M},f)$, determined by the amplitude threshold for detecting individual sources (see also Ref.~\cite{ali-haimoud2026}). This limit allows us to circumvent divergence encountered when the redshift integration is performed down to $z=0$ in a manner consistent with actual observational procedures for defining the genuine gravitational wave background. We show that, if the lowest-order approximation $z_\mathrm{min} \ll 1$ holds, all the higher-order statistics depend on the mass function of SMBHBs only through a single weighted average, $\langle \mathcal{M}^{10/3} \rangle$. This result is derived without assuming any specific functional form for the distribution function of SMBHBs, $d^2n/dzd\log\mathcal{M}$. Our only assumption is that the distribution is separable with respect to the redshift and chirp mass.

As an application of the higher-order statistics defined in this study, we investigate the ratio of cumulants $\kappa_2 / \kappa_1$, which is equivalent to the variance $\mathrm{V}$ divided by the expectation value $\mathrm{E}$. While it has long been known that the total number of sources and the weighted average $\langle \mathcal{M}^{5/3} \rangle$ are degenerate in the expectation value, our ratio $\kappa_2 / \kappa_1$ does not depend on the total number of mergers and reflects the mass function via $\langle \mathcal{M}^{10/3} \rangle / \langle \mathcal{M}^{5/3} \rangle$. While this ratio is more sensitive to the redshift function than the expectation value, this issue will not become a severe limitation until the statistical uncertainty of observed $\mathrm{V}/\mathrm{E}$ reduces to $\lesssim 15$.

We also investigate another combination of higher-order cumulants, $\kappa_2 \kappa_4 / \kappa_3^2$, which is equivalent to the excess kurtosis $\mathrm{K}$ divided by the squared skewness $\mathrm{S}$. Remarkably, this combination does not depend on any physical quantity in the lowest-order approximation and is given by $9/5$ in a model-independent manner at all frequencies. Thus, this combination gives us a consistency relation for the binary-origin hypothesis of the gravitational wave background. We further examine its robustness by relaxing the lowest-order approximation for a wide range of assumed SMBHB distribution functions.
Quantitatively, the deviation is $\lesssim 30\%$ for a gravitational wave amplitude threshold of $h_\mathrm{th} \gtrsim 10^{-14}$, corresponding to the current PTA's sensitivity for individual sources \cite{agazie2023}. 
Since these cumulants are derived assuming that the background consists of a superposition of unresolved gravitational waves from SMBHBs, this hypothesis will be disfavored if the observed $\mathrm{K}/\mathrm{S}^2$ deviates by, e.g., a factor of a few, from 1.8.
Deriving the value of this combination for alternative sources, such as early-Universe processes, remains a subject for future research. While we may reasonably expect that some inequality also holds for other scenarios \cite{2008PhRvD..77b3505S}, it is not necessarily an equality with the same proportionality constant.

In summary, we have explored the potential of higher-order statistics by resolving their divergence. We have demonstrated that they can serve as a viable tool for extracting information about SMBHBs independently of specific population models. This model-independent approach could yield insights into physical processes underlying the final parsec problem through detailed characterization of the SMBHB population. As a future prospect, we plan to develop a practical framework to estimate these higher-order statistics directly from the observed energy density spectrum of the gravitational wave background, taking the observational errors into account. While recent investigations performed under simplified noise models suggest that extracting non-Gaussianity signatures from the statistics of the spectral shape is practically challenging due to dominant pulsar red noise \cite{bernardo2025}, we intend to examine the extent to which these statistics can be estimated by incorporating a detailed treatment of pulsar noise and frequency-dependent analysis, thereby bridging the gap between theoretical calculations of SMBHB population and realistic PTA observations. In particular, the different frequency dependence of higher-order statistics will play a key role in the analysis of the gravitational wave background \cite{lamb2024}.

\begin{acknowledgments}
We thank Atsushi Nishizawa, Hidetoshi Omiya, and Naoki Seto for valuable discussions. This work was supported by Japan Society for the Promotion of Science (JSPS) Grants-in-Aid for Scientific Research (KAKENHI) Grant No. JP26K07062.
\end{acknowledgments}

\appendix

\section{Detailed Derivation of Higher-Order Statistics via Poisson Statistics \label{app:HOS}}

In this Appendix, we provide a detailed derivation of the cumulants, $\kappa_{j}$, of the energy density spectrum of the gravitational wave background, $\Omega_\mathrm{GW}(f)$. We utilize statistical properties of a Poisson distribution. This approach directly links the source population to the observed statistics.

The energy density spectrum $\Omega_\mathrm{GW}(f)$ is the collective signal from discrete sources. To account for the stochastic fluctuations (shot noise), we consider the energy density contained within a frequency bin of width $\Delta \ln f$ centered on $f$. We discretize the parameter space $\theta = \{\log \mathcal{M}, z\}$ into sufficiently small pixels specified by the index $i$. The total contribution $X \equiv \Omega_\mathrm{GW}(f) \Delta \ln f$ in the frequency bin can be expressed as
\begin{equation}
    X = \Omega_\mathrm{GW}(f) \Delta \ln f \approx \sum_{i} \mathcal{A}(\theta_i, f) N_i,
\end{equation}
where $\mathcal{A}(\theta_i, f)$ is the energy density contribution from a single binary and $N_i$ is the number of sources in the $i$th pixel. In this study, we assume that all the sources are located at the center of the frequency bins and neglect the effects of spectral leakage. This approximation is justified in our target range of $10$--$100\,\text{nHz}$, where the impact of leakage from dominant low-frequency components is significantly diminished. Numerical evaluations have confirmed that this approximation maintains good agreement with exact calculations for all but the lowest-frequency bins \cite{lamb2026}.

We assume that $N_i$ follows a Poisson distribution with expectation value $\lambda_i = \langle N_i \rangle$, given by
\begin{equation}
    \langle N_i \rangle = \frac{d^3N}{d\theta_i d\ln f} \Delta \theta_i .
\end{equation}
A defining property of the Poisson distribution is that its cumulant generating function is given by $\Psi(t) = \ln \langle e^{tN} \rangle = \lambda(e^t - 1)$. The ${j}$th order cumulant is obtained by the ${j}$th derivative of the cumulant generating function at $t=0$:
\begin{equation}
    \kappa_{j}[N_i] = \left. \frac{d^{j}}{dt^{j}} \lambda_i (e^t - 1) \right|_{t=0} = \lambda_i.
\end{equation}
This indicates that cumulants of all orders for the number of sources are equal to its expectation value.

Assuming that the sources in different bins are statistically independent, the ${j}$th order cumulant of the sum $X$ is the sum of the individual cumulants. Using the scaling property $\kappa_{j}[aX] = a^{j} \kappa_{j}[X]$, we obtain
\begin{equation}
    \kappa_{j} [X] = \sum_i \mathcal{A}(\theta_i, f)^{j} \kappa_{j} [N_i] = \sum_i \mathcal{A}(\theta_i, f)^{j} \lambda_i.
\end{equation}
In the continuum limit ($\Delta \theta_i \to d\theta$), the summation is replaced by an integral over the population distribution. The ${j}$th order cumulant of the energy density spectrum $\Omega_{\mathrm {GW}}(f)$ is, then, related to $\kappa_{j}[X]$ by
\begin{align}
    &\kappa_{j}[\Omega_\mathrm{GW}(f)] = (\Delta \ln f)^{-{j}} \kappa_{j}[X] \notag \\
    &= (\Delta \ln f)^{1-{j}} \int \mathcal{A}(\theta, f)^{j} \frac{d^3N}{dz d\log \mathcal{M} d\ln f_\mathrm{r}} dz d\log \mathcal{M}.
\end{align}
Substituting the single-source contribution
\begin{align}
    \mathcal{A}(z_i, \mathcal M_i,f) &=\frac{2\pi^2}{3H_0^2} f^2 h^2(z_i, \mathcal{M}_i,f)
\end{align}
and the differential source number $dN$, we obtain the general form Eq.~\eqref{eq: higher-order}, where $h(z_i, \mathcal M_i,f)$ is given by Eq.~\eqref{h2}.
By definition, the first two cumulants, $\kappa_1$ and $\kappa_2$, correspond to the expectation value and the variance, respectively. The third and fourth cumulants, $\kappa_3$ and $\kappa_4$, characterize the skewness and excess kurtosis, respectively. The standardized forms are defined as $\mathrm{S} = \kappa_3 / \kappa_2^{3/2}$ and $\mathrm{K} = \kappa_4 / \kappa_2^2$, providing dimensionless measures of the non-Gaussianity induced by high-mass or nearby sources in the population.

\section{Dependence of $\kappa_{j}$ on $h_\mathrm{th}$ for Models 1 and 5} \label{app:cumulants_and_ratios}

To investigate the validity of the lowest-order approximation in more detail, we present $\kappa_{j}$ as a function of $h_\mathrm{th}$ for Models 1 and 5 with extreme properties in Figs.~\ref{fig:kappa_appVSExact_light} and \ref{fig:kappa_appVSExact_heavy}, respectively. Because $z_\mathrm{min} (\mathcal{M},f)$ for given $h_\mathrm{th}$ is higher for more massive binaries, the approximation degrades for a population dominated by massive sources like Model 5. Considering the fact that higher-order statistics themselves are generally large for such a massive population, we need to be careful about the precision of this approximation if higher-order statistics are detected early in the PTA observations.
\onecolumngrid
\begin{center}
\centering
\includegraphics[width=0.9\textwidth]
{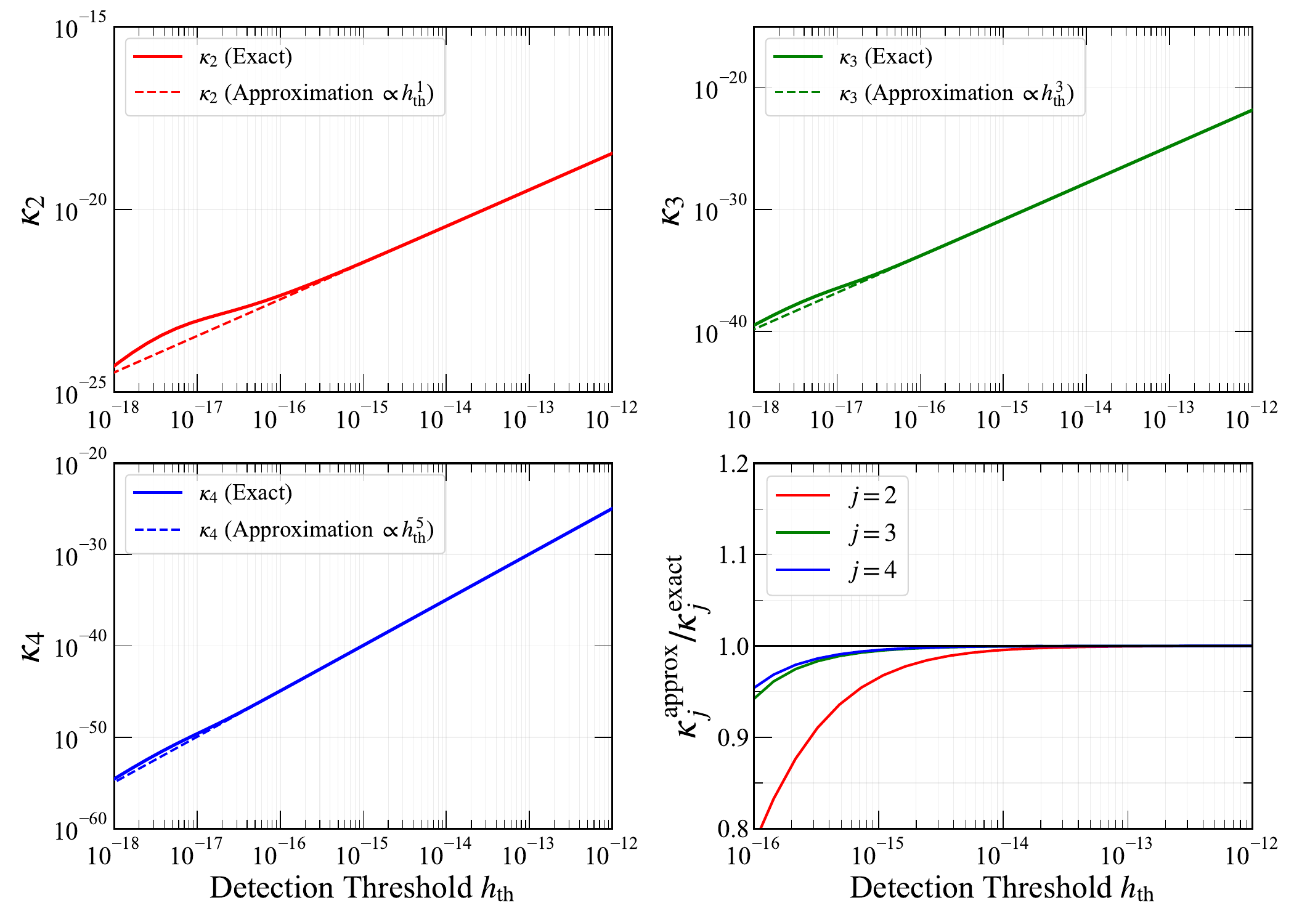}

\captionof{figure}{
Same as Fig.~\ref{fig:kappa_appVSExact} but for Model 1.
}
\label{fig:kappa_appVSExact_light}
\end{center}

\begin{center}
\centering
\includegraphics[width=0.9\textwidth]
{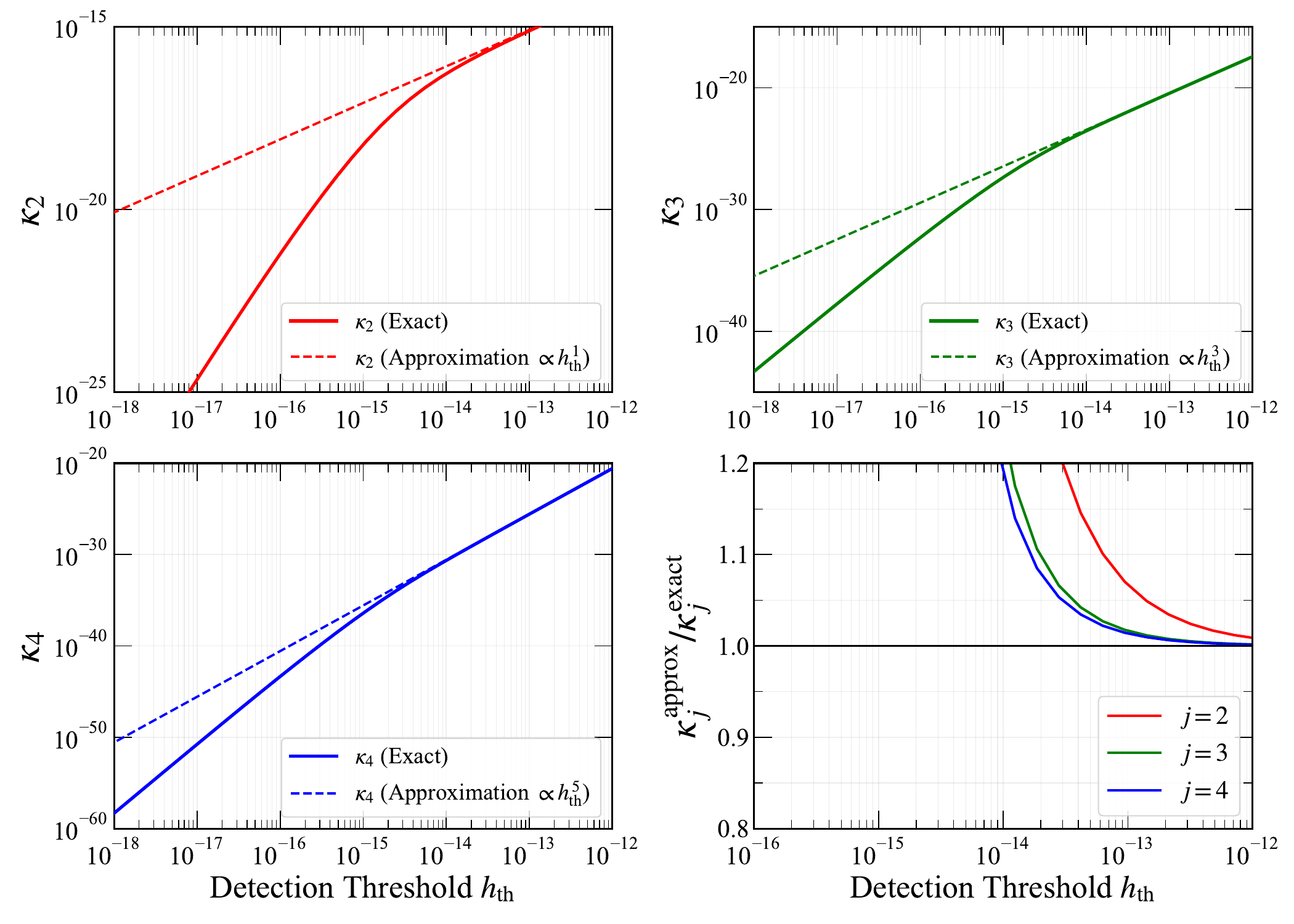}

\captionof{figure}{
Same as Fig.~\ref{fig:kappa_appVSExact} but for Model 5.
}
\label{fig:kappa_appVSExact_heavy}
\end{center}
\onecolumngrid
\vspace{1\baselineskip}
\twocolumngrid
%\bibliography{paper2026}
%apsrev4-2.bst 2019-01-14 (MD) hand-edited version of apsrev4-1.bst
%Control: key (0)
%Control: author (8) initials jnrlst
%Control: editor formatted (1) identically to author
%Control: production of article title (0) allowed
%Control: page (0) single
%Control: year (1) truncated
%Control: production of eprint (0) enabled
%

\end{document}